\documentclass[reprint, longbibliography,noeprint]{revtex4-1}
\usepackage{graphicx}
\usepackage{dcolumn}
\usepackage{bm}
\usepackage{amsmath}
\usepackage{amssymb}
\usepackage{cancel}

\usepackage{hyperref}
\usepackage{xcolor}

\definecolor{webgreen}{rgb}{0,.5,0}
\definecolor{webbrown}{rgb}{.6,0,0}
\definecolor{grigio}{rgb}{.85,.85,.85} 
\definecolor{RoyalBlue}{rgb}{0.0, 0.14, 0.4}
\definecolor{skyblue1}{rgb}{0.45,0.62,0.81}
\definecolor{skyblue2}{rgb}{0.2,0.39,0.64}
\definecolor{skyblue3}{rgb}{0.13,0.29,0.53}
\definecolor{scarlet1}{rgb}{0.93,0.16,0.16}
\definecolor{scarlet2}{rgb}{0.8,0,0}
\definecolor{scarlet3}{rgb}{0.64,0,0}

\definecolor{g}{gray}{0.50}

\hypersetup{%
    colorlinks=true, linktocpage=true, pdfstartpage=1, pdfstartview=FitV,%
    breaklinks=true, pdfpagemode=UseNone, pageanchor=true, pdfpagemode=UseOutlines,%
    plainpages=false, bookmarksnumbered, bookmarksopen=true, bookmarksopenlevel=1,%
    hypertexnames=true, pdfhighlight=/O,
    urlcolor=webbrown, linkcolor=RoyalBlue,  
    pdfsubject={},%
    pdfkeywords={},%
    pdfcreator={pdfLaTeX},%
    pdfproducer={LaTeX REVTeX}%
    }

\begin{document}

\preprint{APS/123-QED}

\title{A [3]-catenane non-autonomous molecular motor model: \\ geometric phase, no-pumping theorem, and energy transduction}

\author{Massimo Bilancioni}
\email{massimo.bilancioni@uni.lu}
\affiliation{Department of Physics and Materials Science, University of Luxembourg, avenue de la Fa\"{i}encerie, Luxembourg City, 1511 G.D.~Luxembourg}
\author{Massimiliano Esposito}
\email{massimiliano.esposito@uni.lu}
\affiliation{Department of Physics and Materials Science, University of Luxembourg, avenue de la Fa\"{i}encerie, Luxembourg City, 1511 G.D.~Luxembourg}
\author{Emanuele Penocchio}
\email{emanuele.penocchio@northwestern.edu}
\affiliation{Department of Physics and Materials Science, University of Luxembourg, avenue de la Fa\"{i}encerie, Luxembourg City, 1511 G.D.~Luxembourg}
\affiliation{Department of Chemistry, Northwestern University, Evanston, 60208 Illinois (United States)}

\date{\today}

\begin{abstract}
We study a model of synthetic molecular motor -- a [3]-catenane consisting of two small
macrocycles mechanically interlocked with a bigger one -- subjected to a time-dependent driving using stochastic thermodynamics.
The model presents nontrivial features due to the two interacting small macrocycles, but is simple enough to be treated analytically in limiting regimes.
Among the results obtained, we find a mapping into an equivalent [2]-catenane that reveals the implications of the no-pumping theorem stating that to generate net motion of the small macrocycles, both energies and barriers need to change.
In the adiabatic limit (slow driving), we fully characterize the motor's dynamics and show that the net motion of the small macrocycles is expressed as a surface integral in parameter space which corrects previous erroneous results. 
We also analyze the performance of the motor subjected to a step-wise driving protocols in absence and in presence of an applied load.
Optimization strategies for generating large currents and maximizing free-energy transduction are proposed.
This simple model provides interesting clues into the working principles of non-autonomous molecular motors and their optimization.
\end{abstract}

\maketitle

\section{Introduction}

Over the last decades, stochastic thermodynamics has developed as a theory describing the energetics of mesoscopic systems driven far from equilibrium ~\cite{Sekimoto2010,vandenbroeck2015,jarzynski2011,seifert2012,rao2018gen,PelitiPigolotti2021}.
It has been used to study systems such as colloidal particles~\cite{raul2017,ciliberto2017,bechhoefer2021}, chemical reaction networks~\cite{nicolis1986,qian2005,seifert2007,rao2018crn}, electronic circuits~\cite{wolpert2020,freitas2020,freitas2021}, and biological molecular motors. 
In this latter case, the quest for a detailed assessment of their thermodynamic performance is particularly important and being actively pursued~\cite{gaspard2006,parrondo2013,seifert2016,sivak2020,leighton2022}.
Surprisingly however, despite the concurrent bloom of artificial molecular motors~\cite{leigh2015,astumian2020pumps,borsley2022singlebond,Electric_motor}, few studies analyzed these systems through the lens of stochastic thermodynamics~\cite{amano2022info,corra2022,penocchio2022bipartite}.
Yet, because the chemistry of these motors is relatively simple~\cite{zerbetto2007,Balzani2008,pezzato2017,amano2022_ch}, elementary models often grasp many key aspects of their kinetics~\cite{astumian2007,astumian2007_adiabatic,astumian2019,albaugh2022,asnicar2022,penocchio2023_diagrams}. 
This makes them ideal case studies for probing the extent to which stochastic thermodynamics can be helpful to deepen our understanding of  their working and suggest ways to design and operate them optimally.
So far, the vast majority of these artificial systems operate non-autonomously, meaning that a directional flow emerges due to a periodic external time-variation of parameters such as electric potential~\cite{pezzato2018,pumm2022,Electric_motor}, light irradiation intensity \cite{theoretical,baroncini2020}, or the concentrations of chemicals~\cite{leigh2004,fletcher2005,branchaud2005,synth_walker,chemical_pulses,feringa2016nonautonomous,feringa2020lactone,zhao2022}.
In the theoretical literature, these non-autonomous systems, often called stochastic pumps, are well understood in the adiabatic limit~\cite{marcus1999,astumian2003,PhysRevE.57.7297,Sinitsyn_2007,Sinitsyn_2009,astumian2007_adiabatic,Parrondo_1998} (i.e., when the parameters are slowly driven) and in the linear regime~\cite{Sokolov_1999,astumian2011,Forastiere_2022} (i.e., for weak perturbations).
Outside these two regimes, a universal no-pumping theorem has been derived~\cite{NPtheorem}, and general comparisons with autonomous molecular motors have been drawn~\cite{mapping,rotskoff}.
However, a comprehensive theory accounting for their behavior in arbitrary regimes is still lacking.
As a result, system-specific studies~\cite{barato,rahav2011} are very valuable for better characterizing the different modes of operations of these non-autonomous molecular motors.
This paper goes precisely in this direction by focusing on non-autonomous catenane-based molecular motors~\cite{theoretical,chemical_pulses,Electric_motor}, i.e., systems composed of two or more mechanically interlocked macrocycles (ring-like molecules).

We based our study on a model of a three-macrocycle catenane motor made of two small macrocycles mechanically interlocked with a bigger one. 
This model was previously introduced in Ref.~\cite{astumian2007_adiabatic,astumian2011}. 
It is simple enough to be treated analytically and, at the same time, presents non-trivial features arising from the presence of the two small interacting macrocycles.
It has been previously studied in the limit of adiabatic operation, where the molecular motor behaves as a reversible pump, and geometric effects reminiscent of the Berry phase in quantum mechanics arise~\cite{astumian2007_adiabatic}.
However, an incorrect formula has been derived to quantify these geometric effects~\cite{astumian2007_adiabatic,astumian2011}.
Here, we correct and further elaborate on it. 
We also find a mapping of the motor dynamics into that of a two-macrocycle catenane that elucidates its relation with the no-pumping theorem.
In addition, we characterize the dynamic and thermodynamic behaviour of the model beyond the adiabatic regime by studying a step-wise driving protocol that mimics how non-autonomous molecular motors are experimentally operated.
We do so both in the absence and presence of a load, finding optimal protocols to maximize specific quantities.
In the first case where there is no output work, we introduce a non-thermodynamic coefficient that measures the motor's performance.
In the second case, we study the output power and the transduction efficiency, and we develop a method for estimating the stopping force.\\

This paper is organized as follows.
We introduce the three-macrocycle catenane motor model in Sec.~\ref{sec:model}, explaining how its non-autonomous operation works (Sec.~\ref{subsec:non-aut}) and discussing its relationship with the no-pumping theorem by leveraging the aforementioned mapping into a two-macrocycle catenane.
In Sec.~\ref{sec:free_dynamics}, we investigate the motor's free dynamics, i.e., its behavior in absence of an applied load.
This includes the adiabatic limit (Sec.~\ref{adiabatic}) and the detailed study of a step-wise driving protocol (Sec.~\ref{freestepprotocol}).
In Sec.~\ref{sec:transduction}, we introduce a load and analyze the ability of the motor to perform free energy transduction~\cite{Brown2020} under the adiabatic~\ref{adiabaticload} and the step-wise driving protocol (Sec.~\ref{loadstepprotocol}), proposing a method to estimate the stopping force in the latter regime (Sec.~\ref{sec:stopping_force}).

In this paper, energy-related quantities will be always expressed in units of $k_BT$ unless otherwise specified.
Furthermore, the subscript ``${cyc}$" will represent the average of the corresponding quantity over a cycle of the driving protocol.

\section{The model}
\label{sec:model}
\begin{figure}[h!]
\centering
\includegraphics[scale=0.41]{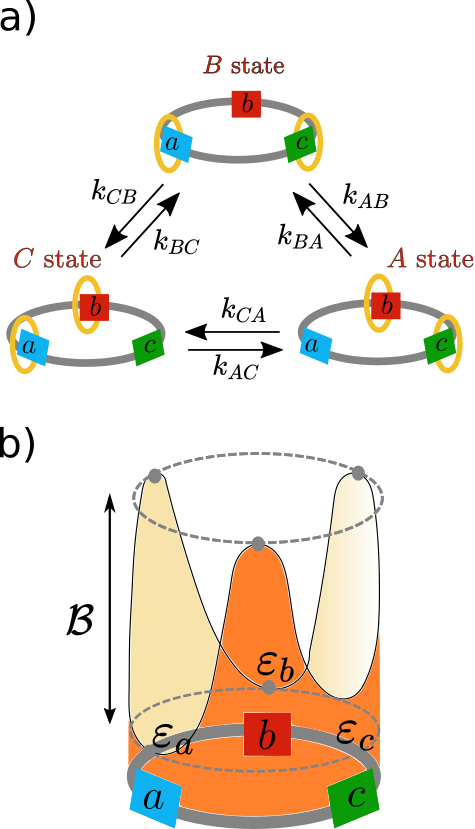}
\caption{The [3]-catenane motor.
\textbf{a)} Chemical reaction network of the molecular motor~\cite{astumian2007_adiabatic,astumian2011}. The [3]-catenane motor comprises two identical small rings (colored in yellow) mechanically interlocked with a larger ring acting as a track for the small rings' shuttling. The track presents three distinguishable stations, denoted $a$, $b$, and $c$, where the small rings sit preferentially due to attractive interactions. Each station can host up to one ring. We therefore consider three possible (meso)states (see Appendix~\ref{app:cg}), denoted $A$, $B$, or $C$ based on which station is unoccupied, connected by three reversible transitions with rate constants $k_{IJ}$. The subscript $IJ$ denotes a transition from state $J$ to state $I$, corresponding to the yellow ring in station $i$ jumping into station $j$.
\textbf{b)} Pictorial representation of the potential free energy surface seen by a ring while shuttling along the track (without taking into account the interaction with the other ring). Each station corresponds to a free energy minimum and the absolute height $\mathcal{B}$ of the barrier between each couple of station is assumed to be the same. Specifically, we depicted a configuration where the $c$ station is less stable compared to the other, favouring state $C$. The exclusion effect preventing the two rings to occupy the same station simultaneously is not represented (see the discussion in the main text).}
\label{fig:model}
\end{figure}

Our case study is a [3]-catenane consisting of two small macrocycles mechanically interlocked with a bigger one (Fig.~\ref{fig:model}a).
The three macrocycles, hereafter denoted as the two \textit{rings} (yellow in Fig.~\ref{fig:model}a) and the \textit{track} (gray in Fig.~\ref{fig:model}a), can move relative to each other.
In the following, we will always refer to the movement of the rings with respect to the track.
The latter hosts three binding sites labeled $a$, $b$, and $c$, namely stations where the rings sit preferentially due to favorable interactions.
The two rings, which we treat as identical, cannot pass one another nor occupy the same station due to steric (i.e., repulsive) interactions between them.

We construct a coarse-grained model of the system in terms of discrete (meso)states: each of these states is the collection of all the possible microscopic configurations in which the two rings occupy a given pair of stations.
This coarse-graining is legitimate when the microscopic dynamics is much faster than the mesoscopic one, as explained in appendix~\ref{app:cg}. 
Overall, due to the identical nature of the rings, the system has three possible states, each labeled by the uppercase letter ($ A,B,C$) corresponding to the unoccupied station (see Fig.~\ref{fig:model}a).
By denoting the binding free energies between the three stations and the rings as $\varepsilon_i$, with $i= a,b,c$ (Fig.~\ref{fig:model}b), the free energies of the states can be expressed \--- up to a constant, which is here set to zero without loss of generality \--- as :
\begin{equation}
E_A = \varepsilon_b +\varepsilon_c \, ; \
E_B = \varepsilon_c +\varepsilon_a \, ; \
E_C = \varepsilon_a +\varepsilon_b \, .
\end{equation}
(See Appendix~\ref{app:cg} for the definition of a mesostate's free energy in this context).
Transitions between states occur when one of the rings jumps in the unique vacant station, affording the chemical reaction network depicted in Fig.~\ref{fig:model}a.
Transitions in which one ring jumps in a station that is occupied by the other ring are prevented by repulsive interactions.
Apart from this exclusion effect, the two rings jump independently from each other.
Assuming that the free energy barrier between any two stations has the same absolute height $\mathcal{B}$ for each transition, the rate constants in the network can be expressed in the Arrhenius form:
\begin{equation}
    \begin{matrix}
 k_{AB}=k_{AC}=\mathcal{A}\,e^{-(\mathcal{B}-\varepsilon_a)}\qquad& \text{(ring in $a$ jumps)}\\
 k_{BC}=k_{BA}=\mathcal{A}\,e^{-(\mathcal{B}-\varepsilon_b)}\qquad &\text{(ring in $b$ jumps)}\\
 k_{CA}=k_{CB}=\mathcal{A}\,e^{-(\mathcal{B}-\varepsilon_c)}\qquad &\text{(ring in $c$ jumps)}\\
    \end{matrix}
    \label{eq:system_rates}
\end{equation}
Note that the activation energy appearing in each rate is the initial energy of the ring that perform the jump. As it should be for thermodynamic consistency, they obey local detailed balance (which in this context corresponds to microscopic reversibility~\cite{amano2022info}):
\begin{equation}
    \frac{k_{IJ}}{k_{JI}}= \exp(E_J-E_I)
    \label{eq:ldb}
\end{equation}

\subsection{Non-autonomous operation}
\label{subsec:non-aut}
The non-autonomous operation of this molecular motor consists of a periodic driving protocol of period $\tau$ that only changes the free energies of the stations without any modification of the barriers' absolute height $\mathcal{B}$.
We assume that any periodic driving protocol defined by a control parameter $\pi(t)$, $\varepsilon_i(t)\equiv\varepsilon_i(\pi(t)) $, is in principle realizable.
This kind of periodic protocols can induce directional flow of the rings around the track.
The  reason can be understood with the help of Fig.~\ref{fig:model}a.
Suppose we start with $\varepsilon_c\gg \varepsilon_a,\varepsilon_b$, so that the system will be with high probability in state $C$.
Then, the free energies of the stations are switched to a new configuration where $\varepsilon_b\gg \varepsilon_c,\varepsilon_a$, so that the ring in $b$ is now in a high energetic station favoring its jump into either $a$ or $b$, but since $a$ is occupied by the other ring, the forward jump into $c$ will be preferred resulting in state $B$.
Then, the free energies of the stations are switched to $\varepsilon_a\gg \varepsilon_b,\varepsilon_c$, so that the ring in $a$ will most likely jump into $b$,  as the ring that previously jumped into $c$ now blocks the backward movement, yielding the state $A$.
After this, the cycle repeats.
We specify that, in the following, we focus on the behavior of the system in the periodic regime, that is, the behavior of the system after many cycles of driving occurred.

The driving protocol leads to time-dependent rates
\begin{equation}k_{IJ}(t) = \mathcal{A}\,e^{-(\mathcal{B}-\varepsilon_i(t))}\end{equation} 
which at any instant satisfy the local detailed balance condition (Eq.~\eqref{eq:ldb}).
As a consequence, the probability distribution evolves according to a  master equation with a time-dependent transition matrix:
\begin{equation} \dot{\boldsymbol{p}} = \mathbb{W}(t)\,\boldsymbol{p}(t) \, , \end{equation}
where
\begin{equation}
\mathbb{W}(t) = \mathcal{A}\,e^{-\mathcal{B}}  \, \mathbb{M}(t)
\label{eq:trans_matrix_chap3}
\end{equation}
and
\begin{equation}
\mathbb{M}(t) = \left(\begin{matrix} -(e^{\varepsilon_b(t)}+e^{\varepsilon_c(t)}) & e^{\varepsilon_a(t)} & e^{\varepsilon_a(t)} \\ e^{\varepsilon_b(t)} & -(e^{\varepsilon_a(t)}+e^{\varepsilon_c(t)}) & e^{\varepsilon_b(t)}\\
e^{\varepsilon_c(t)} & e^{\varepsilon_c(t)} & -(e^{\varepsilon_a(t)}+e^{\varepsilon_b(t)})\end{matrix}\right)
\label{eq:trans_matrix_chap3}
\end{equation}
The eigenvalues of 
 $\mathbb{W}(t)$ are: 
 \begin{itemize} 
     \item  $\lambda_0 (t) = 0$. The corresponding eigenvector is the equilibrium distribution $\boldsymbol{p}^{eq} (t)$ at time $t$ given by:
\begin{equation}
    p_I^{eq} (t) = \frac{e^{\varepsilon_i (t)}}{e^{\varepsilon_a (t)}+e^{\varepsilon_b (t)}+e^{\varepsilon_c (t)}} \, .
    \label{eq:equilibrium}
\end{equation}
     \item $\lambda_{1,2} (t) = -\lambda(t)$ with \begin{equation}
\lambda(t) =  \mathcal{A}\,e^{-\mathcal{B}} \left(e^{\varepsilon_a (t)}+e^{\varepsilon_b (t)}+e^{\varepsilon_c (t)}\right) \, .
\label{eq:ktilde}
\end{equation}
     It can be verified by direct matrix multiplication that the eigenspace corresponding to $\lambda_1$ and $\lambda_2$ is the 2-dimensional subspace of vectors whose components add up to zero.
 \end{itemize}
 The fact that $\mathbb{W}(t)$ has two equal eigenvalues follows from the symmetry of the model and considerably simplifies the master equation that becomes:
 \begin{align}
\dot{\boldsymbol{p}} =& \mathbb{W}(t)\,\left(\boldsymbol{p} (t) - \boldsymbol{p}^{eq} (t)\right) + \underbrace{\mathbb{W}(t)\, \boldsymbol{p}^{eq} (t)}_{= 0} \nonumber \\
=& -\lambda (t)\,\left(\boldsymbol{p} (t) - \boldsymbol{p}^{eq} (t)\right) \, .
    \label{me}
\end{align}
This evolution equation tells us that the time variation of $p_I$ depends only on $p_I$ itself and it relaxes towards the current equilibrium value with rate $\lambda(t)$.\\ 
For simplicity, from now on the time dependence of the free energies $\varepsilon_i$ and of the quantities that depend on them, such as $k_{IJ}$, $\lambda$ and $p_I^{eq}$, will be left implicit in the equations.

\subsection{Relationship with the no-pumping theorem}
\label{sec:no_pumping}
The non-autonomous operation described in the previous section seemingly contradicts the so-called no-pumping theorem~\cite{NPtheorem}, a no-go result stating that in order to produce a directional flow with a cyclic driving protocol, one needs to vary both the energies of the stations and barriers' heights.
However, the no-pumping theorem holds strictly for systems with single particles (as experimentally observed in the version of this model with only one ring ([2]-catenane~\cite{theoretical}) or at most multiple independent particles~\cite{NPmanyparticle}, whereas the system under study comprises two interacting rings.
In a sense, the presence of the two rings can be seen as causing a change in the barriers during the non-autonomous operation of the motor: one of the rings acts in turn as an additional barrier preventing the other ring from moving backward.
We stress that this mechanism only works when the interactions between the two rings are long-ranged, as we implicitly assumed by imposing that the two rings cannot occupy the same station.
If the two rings only interacted locally when in the same station, then an extension of the no-pumping theorem to locally interacting many-particle systems would apply~\cite{NPmanyparticle} and we would not be able to generate directional flow with those kinds of protocols.

\begin{figure}[h!]
\centering
\includegraphics[scale=0.40]{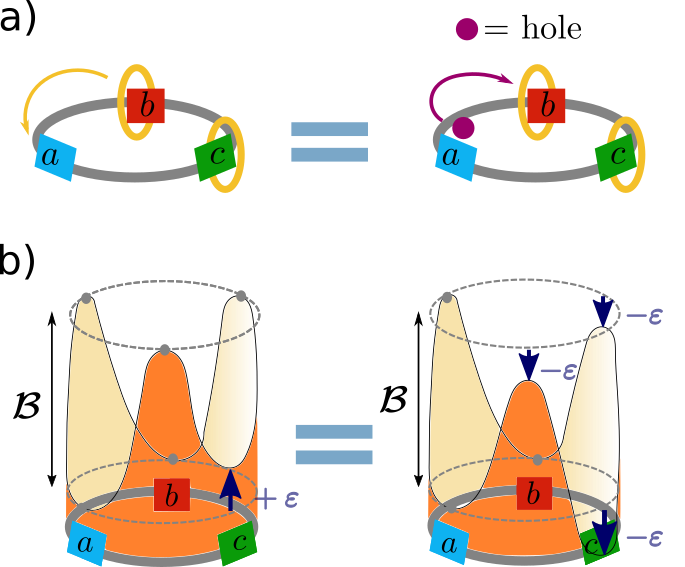}
\caption{ 
\textbf{a)} Two equivalent ways of looking at the $A\to B$ transition. The jump of a small yellow ring from station $b$ to station $a$ (left) can be equivalently looked at as a jump of the \textit{hole} from state $a$ to state $b$ (right).
\textbf{b)} Pictorial representation of the mapping between the [3]-catenane and the [2]-catenane at the level of the potential energy surfaces. Increasing the energy of the $c$-station in the [3]-catenane by $\varepsilon$ (left) is equivalent to lower the energy of the $c$ station and the two adjacent barriers in the equivalent [2]-catenane by the same amount (right).}
\label{fig:mapping}
\end{figure}

To show more rigorously that the way in which directed flow can be induced in the [3]-catenane does not contradict the no-pumping theorem, we now map our system into an equivalent [2]-catenane that produces directional flow in compliance with the no-pumping theorem.
The idea is to describe our system in terms of the vacant station, i.e., the \textit{hole}.
Indeed, the transition $A\to B$ can be alternatively thought of as the hole jumping from $a\to b$ (Fig.~\ref{fig:mapping}a).
We can assign some effective free energies $\varepsilon_i^{h}$ and barriers $\mathcal{B}_{ij}^h$ describing the hole's dynamics so that it exactly reproduces that of the original system.
For this purpose, it is sufficient to choose $\varepsilon_i^{h}$ and $\mathcal{B}_{ij}^h$ such that the transition rates for the hole $k_{IJ}^h$ in the Arrhenius form coincide with the $k_{IJ}$ of the original system:
\begin{equation}k_{IJ}^h = \mathcal{A}\,e^{-(\mathcal{B}_{ij}^h-\varepsilon_j^h)} =\mathcal{A}\,e^{-(\mathcal{B} -\varepsilon_i)}= k_{IJ}\end{equation}
Up to a constant, the correct choice reads:
$$
\begin{gathered}
\varepsilon_{a}^{h}=-\varepsilon_{a} \\
\varepsilon_{b}^{h}=-\varepsilon_{b} \\
\varepsilon_{c}^{h}=-\varepsilon_{c} \\
\mathcal{B}_{ab}^{h}=\mathcal{B}-\varepsilon_{a}-\varepsilon_{b} \\
\mathcal{B}_{bc}^{h}=\mathcal{B}-\varepsilon_{b}-\varepsilon_{c} \\
\mathcal{B}_{ca}^{h}=\mathcal{B}-\varepsilon_{c}-\varepsilon_{a}
\end{gathered}
$$
The effective free energies $\varepsilon_{i}^{h}$ experienced by the hole are the opposite of the ones experienced by the rings: if the hole is in one station, there is no ring, so that the contribution of that station's binding energy is absent.
Finally, by imagining the hole as a single ring interlocked to the track, the mapping is effectively between a [3]-catenane and a [2]-catenane\footnote{An analogous mapping is possible in catenanes with $N$ rings and $N+M$ stations. In that case, the equivalent system would have $M$ rings and same number of stations.}.
A specific example of this mapping is shown in Fig.~\ref{fig:mapping}b, where the two potential energy surfaces experienced by the respective rings are sketched: increasing the free energy of the $c$-station in the [3]-catenane by $\varepsilon$ is equivalent to lower the free energy of the same station and the two adjacent barriers in the [2]-catenane by $\varepsilon$.
As this example shows, a driving protocol which only varies the free energies of the stations in the [3]-catenane corresponds to a driving that varies \emph{both} the free energies and the barriers' heights in the equivalent [2]-catenane.
Crucially, if a driving produces directional flow in the original system, the equivalent driving produces the same flow also in the [2]-catenane, where the no-pumping theorem applies.
Therefore, the generation of current in the [3]-catenane motor is in compliance with the no-pumping theorem.
This explanation is conceptually equivalent to the one already given in \cite{NPtheorem} for such a system, here we made it explicit by leveraging the mapping.

\section{Free dynamics}
\label{sec:free_dynamics}
In this section, we analyze the dynamics and thermodynamics of the [3]-catenane when the free energies of the states are modified according to a periodic driving protocol of the kind described in the previous section.
Interesting quantities to characterize the motor's performance under driving are the average current $J_{cyc}$ generated and the average work $W_{cyc}$ done on the system over a cycle of driving.
The former can be expressed as
\begin{equation}
    J_{cyc} = \frac{\Phi_{cyc}}{\tau}
\end{equation}
where $\Phi_{cyc}$ is the rings' flux over a cycle.
We also introduce a dimensional coefficient of performance (COP) that measures how effective a driving is in producing directional flow:
\begin{equation}
    \text{COP} =\frac{ \Phi_{cyc}}{ W_{cyc}} = \text{number of laps per unit joule spent}
    \label{eq:copdef}
\end{equation}
The idea behind this coefficient is that, at a fixed period, the higher the COP for a certain driving protocol, the less work is required to generate the same rings' flux.

We start by exploring one interesting limiting case, namely the limit of adiabatic (i.e., quasi-static) driving, then we look at one specific type of protocol that models typical experiments \cite{theoretical,chemical_pulses,Electric_motor} and is exactly solvable for every period $\tau$; namely, the step protocol.

\subsection{Adiabatic driving}
\label{adiabatic}
We consider a driving as adiabatic whenever the system's relaxation rate, $\lambda$ in eq.~\eqref{eq:ktilde}, is much faster than the driving protocol, so that the system can be considered to always be in thermodynamic equilibrium with respect to the instantaneous values of stations' free energies.
In this regime, the [3]-catenane behaves as a reversible pump~\cite{PhysRevE.57.7297}, that is, a finite directional flux is generated at the cost of vanishing input work over a period of driving:
\begin{equation}
    \Phi_{cyc} \propto \text{constant} \, ,\quad W_{cyc} \propto 1/\tau \, .
\end{equation}
In addition, the flux $\Phi_{cyc}$ becomes a purely \emph{geometric phase}~\cite{astumian2007_adiabatic,PhysRevE.57.7297} that does not depend on $\tau$ but only on the loop swept by the driving protocol in the space of parameters (i.e., the free energies of the stations).
This property is  analogous to the Berry phase in quantum mechanics \cite{berry}, namely the geometric phase  difference acquired by an eigenstate for a cyclical and adiabatic variation of the Hamiltonian's parameters.

\begin{figure}[h!]
    \centering
    \includegraphics[scale=0.6]{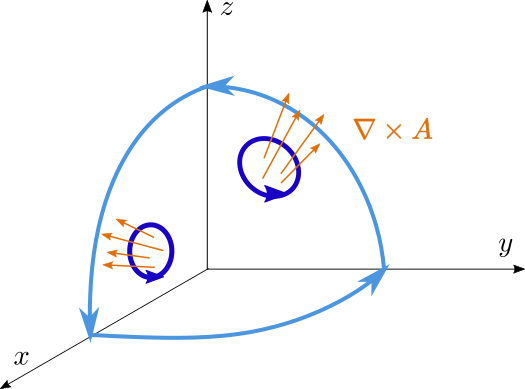}
    \caption{Graphical representation of the geometric phase in terms of the flux of $\nabla \times \boldsymbol{A}$.  The blue closed loops are possible periodic driving protocols in the parameters space ($x,y,z)$. The orange arrows represent the flux across these loops generated by the radial $\nabla \times \boldsymbol{A}$. The big light blue loop corresponds to the driving which collects all of the outgoing flux of $\nabla \times \boldsymbol{A}$. }
    \label{rot}
\end{figure}

In order to formally derive the geometric phase induced in the [3]-catenane by an adiabatic driving protocol, we use eq.~\eqref{eq:ktilde} and \eqref{me} to recast the probability current as:
\begin{align}
    J_{IJ}(t)=& k_{IJ}\,[{p}_J(t)-p_J^{eq}] -k_{JI}\, [p_I(t)-p_I^{eq}]
    =\\ = &-\frac{k_{IJ}\,\dot{p}_J(t) -k_{JI}\,\dot p_I(t)}{\lambda}
    \label{corrente}
\end{align} 
Therefore, the current from  $A\to B$ can be written as: 
\begin{equation}
J_{BA}(t) = \boldsymbol{V}_{BA}\cdot \dot{\boldsymbol{p}}(t)\, ,
\end{equation}
with
\begin{equation}
\quad\boldsymbol{V}_{BA} = \frac{1}{\lambda}(-k_{BA},k_{AB},0) \, .
\end{equation}
The flux over a cycle can then be expressed as
\begin{equation}
    \Phi_{cyc} = \Phi^{BA}_{cyc} =  \int \boldsymbol{V}_{BA}\cdot \dot{\boldsymbol{p}}(t)\,dt \, .
\end{equation}
By implementing the adiabaticity condition (i.e., the probability distribution is at any instant the equilibrium one defined in Eq.~\eqref{eq:equilibrium}), the above equation boils down to
\begin{equation}\Phi_{cyc} = \int \boldsymbol{V}_{BA}\cdot \dot{\boldsymbol{p}}_{eq}\,dt=  \oint \boldsymbol{V}_{BA}\cdot d \boldsymbol{p}_{eq} \, , \label{line}\end{equation}
The last term on the right-hand side is the purely geometric phase.
Note that, since the driving is adiabatic, a finite flux is generated despite the work performed over a cycle is null, yielding a divergent COP.
This is not in violation of the second law of thermodynamics because no work can be extracted out of this finite yet quasi-static directional flux.
To gain intuition, the line integral in Eq.~\eqref{line} can be converted into a more visualizable surface integral by using Stokes theorem.
A preliminary substitution simplifying the next passages is the following:
 \begin{equation}
    \begin{matrix}
        e^{\varepsilon_a} \to x> 0 \\
        e^{\varepsilon_b} \to y>0  \\
        e^{\varepsilon_c} \to z>0 \\
    \end{matrix}
    \label{eq:substitution}
\end{equation}
By evaluating the vector field $\boldsymbol{V}_{BA}$ and $d\boldsymbol{p}^{eq}$ in terms of the new variables $x,\, y,\, z$, the line integral becomes
\begin{equation}\Phi_{cyc}= \oint \boldsymbol{V}_{BA}\cdot d \boldsymbol{p}_{eq} = \oint \boldsymbol{A}(\boldsymbol{r}) \cdot d \boldsymbol{r} \, ,\end{equation}
and by exploiting Stokes theorem we then have
\begin{equation}
    \Phi_{cyc}=\int(\nabla \times \boldsymbol{A}) \cdot d\boldsymbol{S} \, ,
    \label{stokes}
\end{equation}
with 
\begin{equation}
\nabla \times \boldsymbol{A}\,(\boldsymbol{r})=\frac{2}{(x+y+z)^{3}} \, \boldsymbol{r} \, .
\label{rotor}
\end{equation}
Detailed calculations are reported in Appendix~\ref{app:geom_phase}.
We note that the rotor in Eq.~\eqref{rotor} is different from the ones that correspond to Eq.~(6) and~(7) of \cite{astumian2007_adiabatic} or Eq.~(3) and~(4) of \cite{astumian2011}.
The latter rotors result to be nonsymmetric in $x,y$ and $z$, which is inconsistent.
As a matter of fact, the system's symmetry in the three stations $a,b $ and $c$ demands $\Phi_{cyc}$ to be symmetric in $x,y$ and $z$ which, in turn, demands the same symmetry for $\nabla\times\boldsymbol{A}$.
A graphical illustration of the surface integral in Eq.~\eqref{stokes} is given in Fig.~\ref{rot}.
The main advantage of this representation is that, since $\nabla \times \boldsymbol{A}$ is radial, one can easily tell which of the driving protocols give rise to a nonzero average current. 
Furthermore, we can easily identify the protocols maximizing rings' flux as the ones collecting all of the outgoing rotor field  $\nabla \times \boldsymbol{A}$ (light-blue loop in Fig.~\ref{rot}).
For these protocols, the rings complete an entire cycle with unitary probability after each period.
However, they cannot be performed in reality since they would require, for example, that
\begin{equation}\frac xy \to 0 \implies e^{\varepsilon_a-\varepsilon_b}\to \infty \, .\end{equation}
Nevertheless, they  can be approximated arbitrary well, giving a practical method to optimize adiabatic driving protocols in terms of the induced directional flux.

\subsection{Step protocol}
\label{freestepprotocol}
\begin{figure*}[htbp!]
\centering
\includegraphics[scale=0.35]{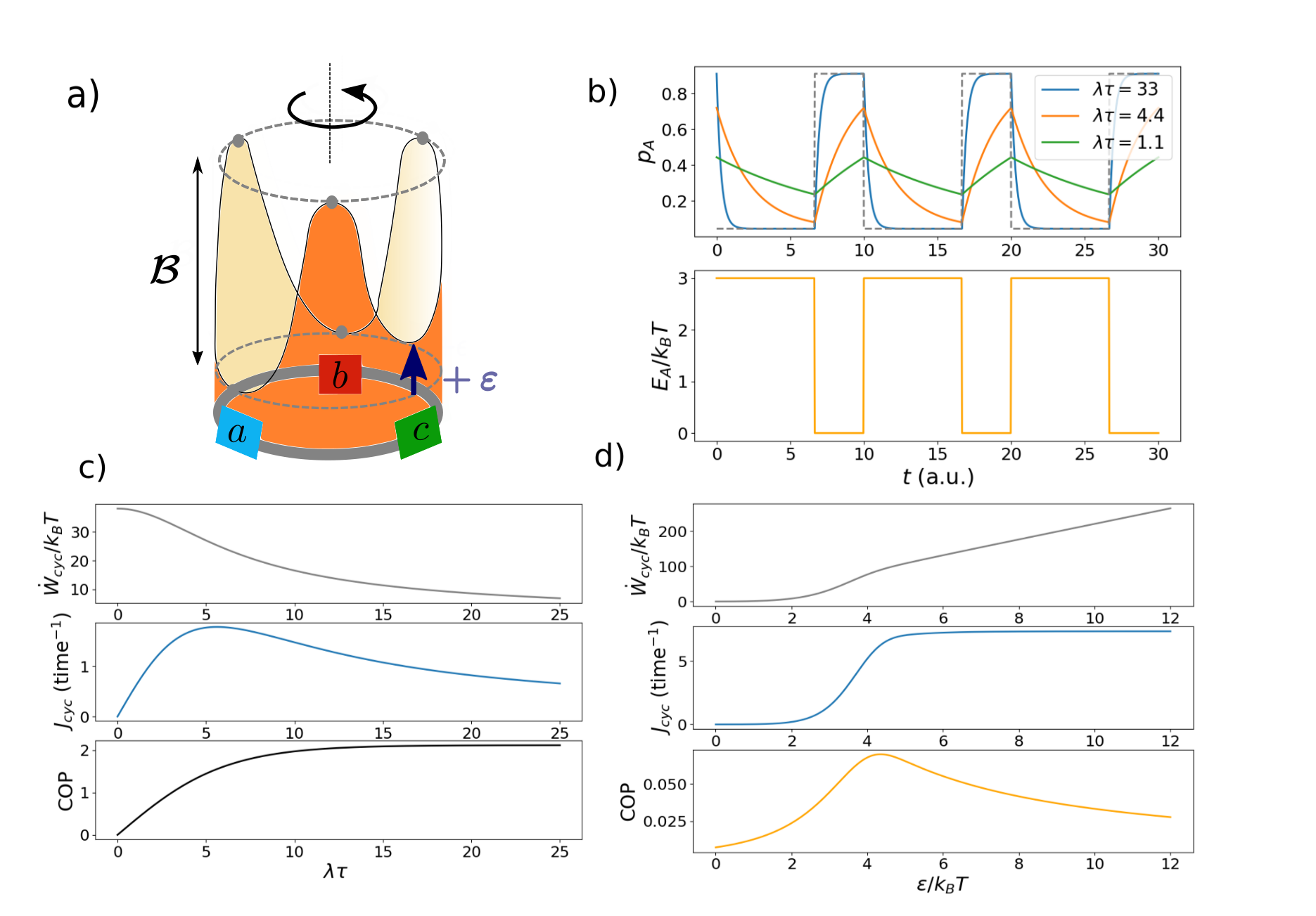}
\caption{Free dynamics under step protocol driving.
\textbf{a)} Each step in the driving can be imagined as a rigid instantaneous rotation of the depicted potential free energy surface by $2\pi/3$.
\textbf{b)} Top: Probability $p_A$ as a function of the time $t$ over three periods $\tau = 10$ a.u.(arbitrary units) for three different relaxation rates $\lambda$.
The dotted line represents the equilibrium value of $p_A$ as a function of time.
Bottom: the energy of the $A$-state during the step protocol: $E_A(t) = \varepsilon_b(t) +\varepsilon_c(t)$.
The modulation energy is $\varepsilon = 3$ $k_B T$.
\textbf{c), d)} Average input power $\dot W_{cyc}/k_B T$ ($\text{time}^{-1}$, Eq.~\eqref{av_input_power}),  current $J_{cyc}$($\mathrm{time}^{-1}$, Eq.~\eqref{current}) and COP ($(k_BT)^{-1}$, Eq.~\eqref{copeq}) as a function of the period $\tau$ with $\varepsilon = 3$ $k_B T$ (c) and of the modulation energy $\varepsilon$ with $\tau= 0.14$ a.u. (d).
The relaxation rate is set to $\lambda =22$ time$^{-1}$ in both cases.}
\label{fig:stepP}
\end{figure*}

In this section, we analyze in detail the step protocol, a type of driving protocol that is exactly solvable and close to what is usually implemented in experiments~\cite{theoretical,Balzani2008,chemical_pulses,Electric_motor}.
At odds with adiabatic protocols in the previous section, in the step protocol, driving is much faster than the system’s relaxation rate so that the probability distribution has no time to change during an external manipulation.
In particular, we focus on a protocol in which, over a period $\tau$, the free energies of the stations as a function of time are:
\begin{equation}
(\varepsilon_a,\varepsilon_b,\varepsilon_c)=
  \begin{cases}
    (0,0,\varepsilon) \quad \text{if} \quad 0<t<\tau/3\\
    (0,\varepsilon,0) \quad \text{if} \quad \tau/3<t<2\tau/3\\
    (\varepsilon,0,0) \quad \text{if} \quad 2\tau/3<t<\tau
  \end{cases}
  \label{stepprot}
\end{equation}
where $\varepsilon>0$ is the \emph{modulation energy} and the steps between an energy configuration and the successive one are assumed to be effectively instantaneous.
As a consequence, the step protocol can never be considered adiabatic, even in the limit of large period $\tau$.\\
A graphic illustration of the step protocol is shown in Fig.~\ref{fig:stepP}a, with the two rings moving clockwise according to the intuitive idea discussed at the beginning of Section~\ref{subsec:non-aut}.

\subsubsection{Solution}
In order to solve for the probability distribution, it is  sufficient to find  $p_A(t)$. Indeed,  $p_B(t)$ and $p_C(t)$ are equal to $p_A(t)$ modulus a temporal translation:
\begin{equation}
    \begin{matrix}
        p_B(t) = p_A(t -\tau/3)\\
        p_C(t) = p_A(t +\tau/3)\\
    \end{matrix}
\end{equation}
By combining Eq.~\eqref{me}, \eqref{eq:ktilde}, and \eqref{eq:equilibrium}, the time evolution of $p_A(t)$ reads:
\begin{equation}
     \dot p_A(t) = k\,e^{\varepsilon_a} -\lambda\,p_A(t) \, ,
     \label{pA}
\end{equation}
where we set $k = \mathcal{A}\, e^{-\mathcal{B}}$.
Note that, in this case, $\lambda$ is constant throughout the step protocol and equal to
 \begin{equation}
 \lambda=k\,\left(e^{\varepsilon_a}+e^{\varepsilon_b}+e^{\varepsilon_c}\right)=k\,(2+ e^{\varepsilon}) \, .
 \end{equation} 
The solution of Eq.~\eqref{pA} reads:
\begin{widetext}
\begin{equation}
    p_A(t) =
  \begin{cases}
    e^{-\lambda\, t}\,p_A(0) +\frac{k}{\lambda}\,\,\left(1-e^{-\lambda\, t}\right)       & \quad \text{if } 0\le t < \frac{2\tau}{3}\\
    e^{-\lambda\, t}\,p_A(0) +\frac{k}{\lambda}\,\left(e^{\varepsilon} + (1-e^{\varepsilon})\,e^{-\lambda\, (t-2\tau/3)}-e^{-\lambda\, t}\right) & \quad \text{if } \frac{2\tau}{3}\le t<\tau 
  \end{cases}
  \label{eq:distribution}
\end{equation}
\end{widetext}
with
\begin{equation}
    p_A(0)=\frac{1}{2+e^{\varepsilon}}
    \, \frac{e^{\varepsilon}+x+x^2}{1+x+x^2}\qquad 
    x =\exp\left(-\frac{\lambda\, \tau}{3}\right) \, .
\end{equation}
In Fig.~\ref{fig:stepP}b, $p_A(t)$ is plotted for three different values of $\lambda\, \tau$ and contrasted with the energy of state A during the step protocol. 
As expected, $p_A(t)$ peaks whenever $E_A$ is minimum.  However, we can appreciate how $p_A(t)$ straighten out when $\lambda \tau$ gets smaller, as the system has less time to relax between one step and the other.

\subsubsection{Work}
The total work done in a cycle can be calculated, according to stochastic thermodynamics \cite{PelitiPigolotti2021,seifert2012,vandenbroeck2015}, as 
\begin{equation}
    W_{cyc}=\sum_I\int_0^\tau p_I \dot E_I \, dt,
    \label{eq:work}
\end{equation}
Calculations, reported in Appendix~\ref{app:work}, yield the following expression:
\begin{equation}
  W_{cyc}= 3\,\varepsilon\,\frac{1-e^{-\varepsilon}}{1+2e^{-\varepsilon}}\,\frac{1-x^2}{1+x+x^2}\qquad  x =\exp\left(-\frac{\lambda\, \tau}{3}\right) \, .
   \label{work}
\end{equation}
The average input power supplied by the driving is then
\begin{equation}
    \dot W_{cyc} = \frac{W_{cyc}}{\tau} \, ,
    \label{av_input_power}
\end{equation}
which is plotted in the upper graphs of Figs.~\ref{fig:stepP}c-d as a function of the period $\tau$ and the modulation energy $\varepsilon$. The average input power $\dot{W}_{cyc}$ decreases monotonically as a function of the period $\tau$ (Fig.~\ref{fig:stepP}c), as the work is delivered over a longer time. It also increases monotonically with the modulation energy $\varepsilon$ (Fig.~\ref{fig:stepP}d) due to the higher work required to change the free energies of the stations.

\subsubsection{Current}
The average current in a cycle can be found by integrating, over a period, the current through an arbitrary edge of the network in Fig.~\ref{fig:model}a, e.g., the $A\to B$ edge:
\begin{equation}
    J_{cyc}\ = -\frac{1}{\tau}\int_0^\tau \,J_{BA}(t')\,dt' \, ,
    \label{eq:current}
\end{equation} 
 the negative sign comes from the fact that $J_{BA}$ represents the flow of the \textit{hole} (the A state is the one in which station $a$ is unoccupied), which is opposite to the flow of the two rings (Fig.~\ref{fig:mapping}a).
After the calculations reported in Appendix~\ref{app:current}, the expression of the current boils down to:
\begin{equation} 
    J_{cyc}=\frac{1}{\tau}\,\frac{(1-e^{-\varepsilon})^2}{(1+2\,e^{-\varepsilon})^2}\,\frac{(1-x)^3}{1-x^3} \qquad
     x =\exp\left(-\frac{\lambda\, \tau}{3}\right) \, ,
    \label{current}
\end{equation}
which is plotted in the middle graphs of Figs.~\ref{fig:stepP}c-d as a function of the period $\tau$ and the modulation energy $\varepsilon$.
Interestingly, there is an optimal period $\tau$ for the driving protocol that maximizes the output current $J_{cyc}$ (Fig.~\ref{fig:stepP}c). This optimal period corresponds to the best trade-off between a too fast driving, which does not allow the system to relax between one step and the other, and a too slow driving, which waits too much time after the system relaxed.
Furthermore, when the modulation energy $\varepsilon$ is below a certain threshold, the output current is almost null, and it reaches a plateau very quickly when the threshold is passed (Fig.~\ref{fig:stepP}d).
This on/off behavior can be explained by the fact that the modulation energy $\varepsilon$ must be high enough to beat thermal fluctuations and make the rings able to discriminate the least energetic stations.
At the same time, $e^{-\varepsilon}$ quickly becomes negligible in Eq.~\eqref{current}, yielding a constant current.

\subsubsection{Coefficient of performance}
The analytical expression for the COP as defined in Eq.~\eqref{eq:copdef} can be derived from Eqs.~\eqref{work} and~\eqref{current}:
  \begin{equation}
 \text{COP} = \frac{J_{cyc}\, \tau}{W_{cyc}}= \frac{1}{3\varepsilon}\frac{1-e^{-\varepsilon}}{1+2e^{-\varepsilon}}\frac{1-x}{1+x}\qquad x = \exp\left(-\lambda\,\tau/3\right) \, .
     \label{copeq}
  \end{equation}
The COP is plotted in the bottom graphs of Figs.~\ref{fig:stepP}c-d as a function of the period $\tau$ and the modulation energy $\varepsilon$.
As it does not take into account the speed of operation but only how efficiently a directional flux is produced, it is maximized for long periods (Fig.~\ref{fig:stepP}c).
Furthermore, contrary to the case of adiabatic driving, the COP remains finite in the limit of large $\tau$.
This is due to the fact that, as previously mentioned, the step protocol is never adiabatic, even for large $\tau$.
Finally, the bottom plot of Fig.~\ref{fig:stepP}d reveals the presence of an optimal modulation energy $\varepsilon$ maximizing the COP for a fixed period $\tau$.
The presence of such a maximum can be explained intuitively by considering that, on the one hand, a too high modulation energy $\varepsilon$ is counterproductive because more work is performed without increasing the current; on the other hand, a too small modulation energy $\varepsilon$ does not sufficiently promote forward transitions over backward ones.

\section{Dynamics with applied load}
\label{sec:transduction}
\begin{figure*}[htbp!]
\centering
\includegraphics[scale=0.35]{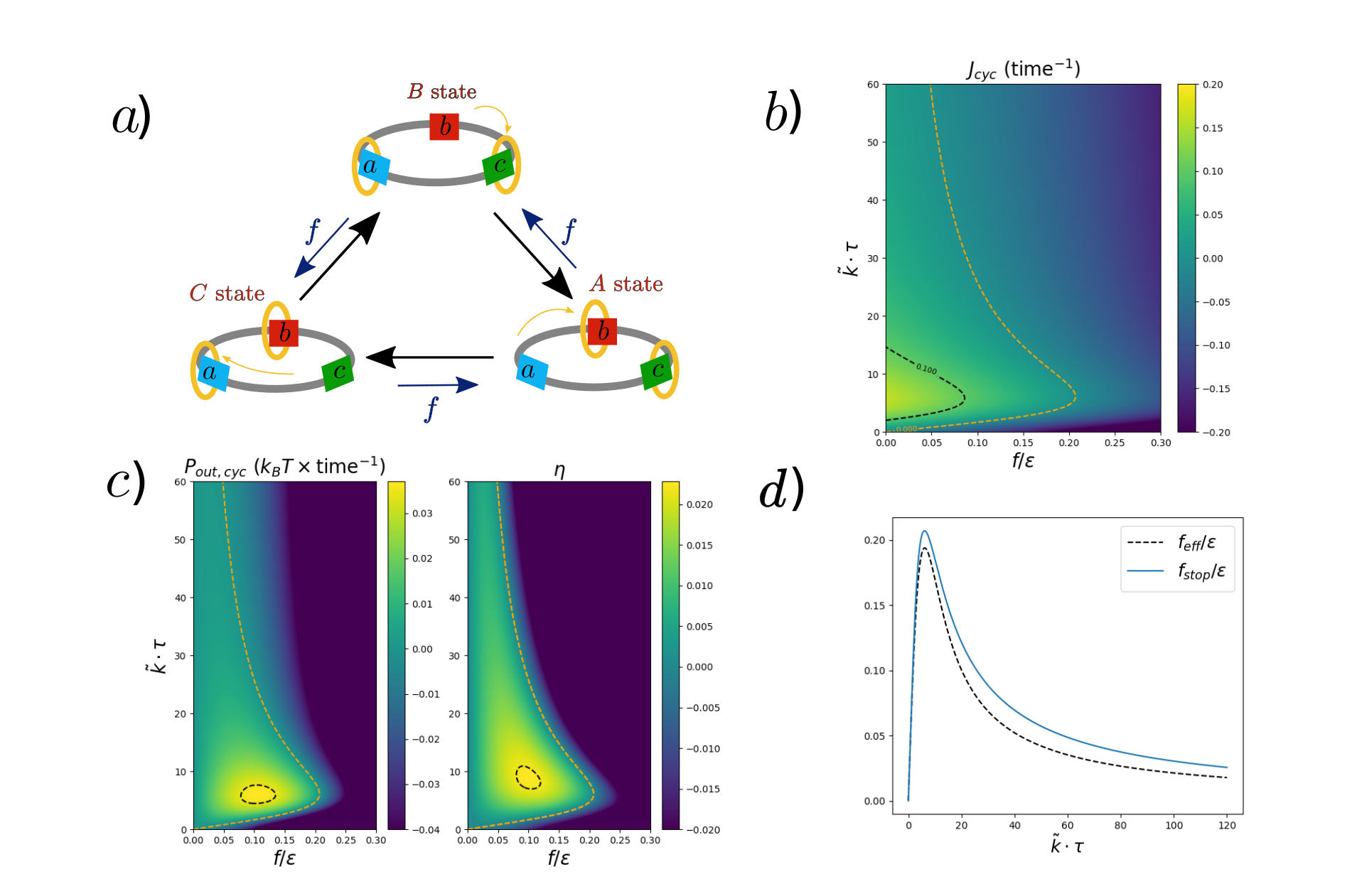}
\caption{Dynamics with applied load under step protocol driving.
\textbf{a)} [3]-catenane motor with an applied load. The load is modelled as an opposite force pushing each ring anticlockwise.
\textbf{b), c)} Average current $J_{cyc}$, output power $P_{out}$, and efficiency $\eta$ as a function of the period $\tau$ and the applied force $f$.
In each plot, the dotted yellow line delimits the region in which work is performed by the motor, i.e., the $J_{cyc}>0$ region. The dashed black lines correspond to the regions in which $J_{cyc}>0.1$ b), $P_{out}>0.33$ and $\eta >0.21$ c).
The modulation energy $\varepsilon = 1.39$ $k_B T$ and relaxation rate $\lambda=6$ time$^{-1}$) are fixed.
\textbf{d)} Comparison of the numerical stopping force $f_\text{stop}$ with the estimate $f_\text{eff}$ as a function of the period $\tau$.
This was done for the step protocol with modulation energy $\varepsilon = 1.39$ $k_B T$ and relaxation rate $\lambda=6$ (time$^{-1}$).}
\label{fig:load}
\end{figure*}

In this section, we study the non-autonomous operation of the [3]-catenane motor under driving and in the presence of a load.
The latter is modeled as an opposing force $f$ applied to each transition (see Fig.~\ref{fig:load}a) so that the total force applied to the three-state motor is $3f$.
Here, we are interested in quantifying the output power and the efficiency with which the input work is converted into the output work done against the force due to the rings moving ahead.
If the forward current is $J_{cyc}$, the average output power delivered by the motor will be
 \begin{equation}
      P_{out} =  3f\, J_{cyc} \, ,
     \label{pout}
 \end{equation}
 and the efficiency 
\begin{equation}
    \eta = \frac{ P_{out}}{\dot W_{in}} \, ,
    \label{efficiency}
\end{equation}
where $\dot W_{in}$ is the average work per unit of time performed by driving the molecular motor.

For any transitions, for instance, the $ A \to B$ transition, the local detailed-balance condition requires now
\begin{equation}
\frac{k_{AB}}{k_{BA}} = e^{E_B-E_A -f} \, .
\end{equation}
In general, there are no constraints on how the force $f$ modifies each rate constant, this will depend on the specific system at study.
Here, we assume that $f$ only modifies the backward rates meaning for example:
\begin{equation}k_{AB} = \mathcal{A}\,e^{-(\mathcal{B}-\varepsilon_a)}, \quad k_{BA} =\mathcal{A}\,e^{-(\mathcal{B}-\varepsilon_b) +f}  \end{equation}
According to this choice, the transition matrix $\mathbb{W}$ in  Eq.~\eqref{eq:trans_matrix_chap3} is replaced by:
\begin{equation}
\mathbb{W}^f(t) = \mathcal{A}\,e^{-\mathcal{B}}  \,\mathbb{M}^f(t) \, ,
\end{equation}
with
\begin{equation}
\mathbb{M}^f(t) =  \left(\begin{matrix} -(e^{(\varepsilon_b+f)}+e^{\varepsilon_c}) & e^{\varepsilon_a} & e^{(\varepsilon_a+f)} \\ e^{(\varepsilon_b +f)} & -(e^{\varepsilon_a}+e^{(\varepsilon_c+f)}) & e^{\varepsilon_b}\\
e^{\varepsilon_c} & e^{(\varepsilon_c+f)} & -(e^{(\varepsilon_a+f)}+e^{\varepsilon_b})\end{matrix}\right) \, .
\end{equation}
Contrary to $\mathbb{W}$, $\mathbb{W}^f$ does not have two identical eigenvalues.
This prevents us from finding a simple solution to the master equation as done in the previous sections and makes numerics necessary to obtain quantitative results.

\subsection{Adiabatic driving with applied load}
\label{adiabaticload}
When subjected to a load, the regime of adiabatic driving is not interesting because the output work vanishes. As a matter of fact, to produce output work, we must have $J_{cyc}>0$. This means that the contribution to $J_{cyc}$ coming from the driving must prevail over the negative contribution arising from the opposing force. The former scales $\propto {1}/{\tau}$ from Sec.~\ref{adiabatic}, while the latter scales $ \propto f$ for small forces. Therefore, in this regime, to observe a current in the direction opposite to the force, the latter must scale as
\begin{equation}
    f\propto \frac{1}{\tau}\, ,
\end{equation}
Since the output power is proportional to the product of the current and the force, the above scaling  implies 
\begin{equation}
    P_{out} \propto \frac{1}{\tau^2} \, ,
\end{equation}
which translates into a vanishing output work over a period:
\begin{equation}
    W_{cyc}\propto \frac{1}{\tau}\, .
\end{equation}

\subsection{Step protocol with applied load}
\label{loadstepprotocol}
For the step protocol, we report the results obtained by lengthy analytical calculations
done in Mathematica~\cite{Mathematica}.
To make such calculations feasible, the modulation energy $\varepsilon$ of the step protocol and $k = \mathcal{A}\, e^{-\mathcal{B}}$  were fixed to specific values ($\varepsilon = \log 4$, $k =  1$), and we only kept track of the analytical dependencies of the motor performance on the period $\tau$ and the force $f$.
In Fig.~\ref{fig:load}b, the current $J_{cyc}$ is plotted as a function of these two parameters, with the yellow dotted line delimitating the area of parameters space in which $J_{cyc}>0$, that is, where we can produce output work.
For any given $\tau$, there is a value of the force, called stopping force $f_{\text{stop}}$, above which the current becomes negative.
We also see that, at fixed $f$, there is a finite range of intermediate periods in which $J_{cyc}>0$.
The reason is that for small and large periods the forward current produced by the driving tends to zero (see Fig.~\ref{fig:stepP}c), and thus the backward current generated by the opposite force dominates.
In Fig.~\ref{fig:load}c, we plotted $P_{out}$ and the efficiency $\eta$ as a function of both $\tau$ and $f$.
The plots show that there is good overlap of the regions in which they are maximum, a feature that can emerge only when systems are operated far from the linear regime \cite{PhysRevLett.105.150603,Penocchio2019}.
Our analysis allows one to identify regions of good tradeoff between power and efﬁciency for the non-autonomously operated [3]-catenane motor.
 
\subsection{Estimating the stopping force}
\label{sec:stopping_force}
As noticed above, for any period $\tau$, a stopping force $f_{\text{stop}}$ can be identified such that the motor is stalled (i.e., $J_{cyc} = 0$).
Knowing the value of $f_{\text{stop}}$ can be useful, as it sets an upper bound to the ability of the motor to perform work against a force under non-autonomous driving.
However, as we discussed, while the free dynamics of the motor can be easily solved, the dynamics in the presence of a load has much greater analytical complications.
Therefore, the question we ask in this section is: can we estimate  $f_{\text{stop}}$ from the free dynamics studied in Sec.~\ref{sec:free_dynamics}?
We start from the intuition that the greater the current pumped by a certain driving in absence of any load, the greater $f_{\text{stop}}$ will be for that driving protocol.
We then notice that the exact stopping force would be easily deducible from the free dynamics if our molecular motor were autonomously driven.  In that case, the stopping force would be the log-ratio of the product of forward and backward autonomous rates.
Based on these considerations, we can proceed as follows: (i) starting from the free dynamics of the non-autonomous [3]-catenane motor, we construct an  ancillary autonomous dynamics~\cite{rotskoff} that, at steady state, has the same probability distribution, current and traffic ($t_{IJ} = k_{IJ}p_J+k_{JI}p_I$) as the original one; (ii) we compute the driving affinity of the ancillary dynamics and take it as an estimate ($f_{\text{eff}}$) for the stopping force of the non-autonomous dynamics; finally, (iii) we compare the estimated $f_{\text{eff}}$ with the real stopping force $f_{\text{stop}}$ in the regimes that we solved.
The construction in point (i) can be easily carried out.
Indeed, it is enough to choose the  rates for the autonomous ancillary dynamics in the following way
\begin{equation}
    k_{IJ}^\text{Aut}  = \langle k_{IJ}  p_J \rangle/\langle p_J\rangle \, ,
    \label{rate}
\end{equation}
where, on the right-hand side, the brackets denote the average over a period in the original non-autonomous dynamics.
This choice ensures that the average probability distribution, current and traffic of the non-autonomous molecular motor are exactly reproduced by the ancillary dynamics:
\begin{equation} p_I^\text{Aut} = \langle p_I\rangle \quad J_{IJ}^\text{Aut} = \langle J_{IJ}\rangle \quad  t_{IJ}^\text{Aut} = \langle  t_{IJ}\rangle\end{equation}
Intuitively, the ancillary autonomous dynamics represents a stroboscopic version of the non-autonomous one where just the average motion over a period is observed.
Moving to point (ii), we estimate the stopping force as the driving affinity of the ancillary dynamics:
\begin{equation}
    f_{\text{eff}} = \log\left(\frac{\prod_\rho k_{+\rho}^\text{Aut}}{\prod_\rho k_{-\rho}^\text{Aut}}\right) \, ,
    \label{fstopping}
\end{equation}
where $k_{+\rho}^\text{Aut},k_{-\rho}^\text{Aut}$ represent the forward and backward rates of the $\rho$ transition, respectively.
Finally, following point (iii), in Fig.~\ref{fig:load}d we compare $f_{\text{eff}}$ with the real stopping force  $f_{\text{stop}}$ as a function of the period $\tau$ of the step protocol.
We find that $f_{\text{eff}}$ is a lower bound for $f_{\text{stop}}$ that qualitatively reproduces its behavior.
In the limit $\tau\to 0$, the two curves converge. 
The reason is that in this limit, the driving becomes so fast that the opposite force effectively only perceives its average effect, which is exactly reproduced by the ancillary dynamics.\\
We conclude this section with some remarks.
The same procedure that we applied to the step protocol can be followed for any other driving.
However, the fact that our estimate lower bounds the exact stopping force has only been tested for the step protocol and for a specific value of the modulation energy ($\varepsilon = 4$), it is not obvious if it holds in general.
Moreover, there are other similar ways of estimating the stopping force from the free dynamics, the one adopted here ensures that, in the limit $\tau \to 0$, the estimate becomes exact.

\section{Conclusions}
Artificial non-autonomous molecular motors are currently in the spotlight of the experimental community working on molecular machines~\cite{zhao2022,pumm2022,Electric_motor}.
In this paper, we applied the tools of stochastic thermodynamics to build a comprehensive understanding of the dynamics and thermodynamics of a simple model epitomizing the functional elements of catenane-based non-autonomous synthetic motors~\cite{leigh2004,chemical_pulses,Electric_motor}.
Our main results can be summarized as follows.
First, we discussed how the current generation in a [3]-catenane relates to the no-pumping theorem~\cite{NPtheorem} leveraging a mapping with an equivalent [2]-catenane.
Second, we corrected and further elaborated on a previously derived formula for the adiabatic limit's geometric flux~\cite{astumian2007_adiabatic,astumian2011}.
Finally, we went beyond the linear and adiabatic regime by studying a step-wise driving protocol that resembles those used in experiments.
We did so by solving for the molecular motor's behavior both in the absence and presence of a load.
In the former case, we quantified its performance by introducing an additional non-thermodynamic coefficient, which we denoted as COP.
In the latter case, we studied the transduction efficiency, the output power and the stopping force. In both situations, we found optimal protocols that maximize specific molecular motor performance quantifiers.
Our study will help the experimental community develop a more in-depth intuition on optimally designing and operating non-autonomous molecular motors.

\section{Acknowledgements}
This is research was supported by AFR PhD grant 15749869 and project ChemComplex (C21/MS/16356329), both funded by the FNR (Luxembourg).
For open access, the author has applied a Creative Commons Attribution 4.0 International (CC BY 4.0) license to any Author Accepted Manuscript version arising from this submission.
\section{Data availability}
The data that support the findings of this study are available from the corresponding author upon reasonable request.
\appendix
\section{Coarse-graining}
\label{app:cg}

In Sec.~\ref{sec:model}, we introduced a coarse-grained model of the molecular motor in terms of discrete mesostates.
Each of these mesostates is a collection of all the different microscopic configurations in which the rings occupy a given pair of stations.
In this appendix, we explain when such an effective description works and define the free energies of mesostates \cite{PelitiPigolotti2021}.

A reliable coarse-graining of a physical system is possible whenever different sets of microscopic configurations (microstates) can be collected into mesostates such that the equilibration at the level of the mesostates is much slower than that of the microstates inside each mesostate.
Under this condition, the microscopic configurations collected into a mesostate can be considered to always be in thermodynamic equilibrium while focusing on the dynamics at the level of the mesostates.
In our coarse-grained treatment of the [3]-catenane motor, we therefore assumed that jumps between stations occur on a much slower time scale than the microscopic dynamics inside the stations.\\
In this scenario, the occupation probability at equilibrium of a given mesostate $I$ in terms of the microscopic states is given by
\begin{equation}
p_i^{eq} = \sum_{\xi\in i}p_\xi^{eq} \propto \sum_{\xi\in I} e^{-\varepsilon_\xi/k_B T} = e^{-E_I/k_B T} \, ,
\end{equation}
where the index $\xi$ runs over all the microstates in the mesostate $I$, $\varepsilon_\xi$ labels the energy of microstate $\xi$, and $E_i$ is precisely the free energy of the mesostate $I$ defined as
\begin{equation}
    E_i=- k_BT\, \log \left(\sum_{\xi\in i} e^{-\varepsilon_\xi/k_BT}\right) \, .
\end{equation}

\section{Geometric phase calculations}
\label{app:geom_phase}
The substitution 
 \begin{equation}
    \begin{matrix}
        e^{\varepsilon_a} \to x> 0 \\
        e^{\varepsilon_b} \to y>0  \\
        e^{\varepsilon_c} \to z>0 \\
    \end{matrix}
\end{equation}
leads to 
\begin{equation}
    p_A^{eq}= \frac{x}{x+y+z}\,, \quad p_B^{eq}= \frac{y}{x+y+z} \, ,
\end{equation}
and 
 \begin{equation}
    \begin{matrix}
      \lambda = k \,(x+y+z) \, ,\\
      \\
    k_{AB} = k\, x \, , \quad \text{and}\quad k_{BA} = k\, y \, ,
    \end{matrix}
\end{equation}
where we set $k = \mathcal{A}\, e^{-\mathcal{B}}$.
By evaluating $\boldsymbol{V}_{BA}$ and $d\boldsymbol{p}^{eq}$ in terms of the new variables $x,\, y$ and $ z$ we get
\begin{equation}
    \Phi_{cyc}  = \oint \boldsymbol{A}(\boldsymbol{r}) \cdot d \boldsymbol{r}
\end{equation}
with 
\begin{equation}\boldsymbol{A}(\boldsymbol{r})=\frac{(-y,x,0)}{(x+y+z)^{2}}  \end{equation}
The calculation of 
$\nabla\times\boldsymbol{A}$ yields
\begin{equation}\nabla \times \boldsymbol{A}=\frac{2}{(x+y+z)^{3}} \, \vec{r}\end{equation}

\section{Step protocol}
\subsection{Calculation of the work}
\label{app:work}
From symmetry arguments the work in eq.~\eqref{eq:work} is equal to:
\begin{equation}
    W_{cyc} = 3\int_0^\tau p_A \dot E_A dt
\end{equation} 
during the step protocol, the energy of the state $A$ is :
\begin{equation}
    E_A =\varepsilon_b + \varepsilon_c= 
  \begin{cases}
       \varepsilon   & \quad \text{if } 0\le t < \frac{2\tau}{3}\\
  0  & \quad \text{if } \frac{2\tau}{3}\le t<\tau 
  \end{cases}
\end{equation}
therefore $W_{cyc}= 3 \,\varepsilon\,( p_A(\tau)-p_A(2\tau/3)) $ and using eq.~\eqref{eq:distribution} for $p_A(t)$ we get
\begin{equation}
          W_{cyc}= 3\,\varepsilon\,\frac{1-e^{-\varepsilon}}{1+2e^{-\varepsilon}}\,\frac{1-x^2}{1+x+x^2}\qquad  x =\exp\left(-\frac{\lambda \,\tau}{3}\right)
\end{equation}
\subsection{Calculation of  the current}
\label{app:current}
From eq.~\eqref{eq:current} the average current in a cycle is:
\begin{equation}
\begin{matrix}
    J_{cyc}  =- \frac{1}{\tau}\int_0^\tau \,k_{BA}\,p_A(t')-k_{AB}\,p_B(t')\,dt'\\
    \\
    =- \frac{k}{\tau}\int_0^\tau \,e^{\varepsilon_b}p_A(t')-e^{\varepsilon_a}p_B(t')\,dt'
    \end{matrix}
\end{equation}
keeping in mind that $p_B(t) = p_A(t+\tau/3)$ and the behavior of $\varepsilon_a$ and $\varepsilon_b$ in the step protocol, we get:
\begin{equation}
    J_{cyc} =\frac{k}{\tau}(e^\varepsilon-1)\left[\int_0^{\tau/3}p_A(t)\,dt- \int_{\tau/3}^{2\tau/3}p_A(t)\,dt\right]
\end{equation}
using eq.~\eqref{eq:distribution} for $p_A(t)$ we finally have:
\begin{equation}
      J_{cyc}=\frac{1}{\tau}\frac{(1-e^{-\varepsilon})^2}{(1+2\,e^{-\varepsilon})^2}\frac{(1-x)^3}{1-x^3}\qquad  x =\exp\left(-\frac{\lambda\, \tau}{3}\right)
\end{equation}
\bibliography{biblio}

\end{document}